# Sequencing Participatory Action Research and *i** Modeling Framework in Capturing Multiple Roles Requirements

Siti Nurul Hayatie Ishak, Ariza Nordin

Faculty of Computer and Mathematical Sciences
MARA University of Technology
Shah Alam, Selangor, Malaysia
{nurulishak14@yahoo.com}

**Abstract-** This paper presents the conceptual framework for sequencing of Participatory Action Research (PAR) methodology with the implementation of i* modeling framework in capturing multiple roles requirements. There are multiple roles involved in the development of information system, thus it involves with difference users requirements and preferences, context as well as the demands which become a challenge in development of system. This is due to these roles where information of the project monitoring is perceived in accordance to their role and domain. In the development of information systems, requirement engineering is a vital methodology. Requirement engineering (RE) consists of several phases which elicitation is a crucial phase in RE since it requires researcher to gather the requirement from the users. Methods of eliciting requirements are now more co-operative. Based on the preliminary study of construction-based in Malaysia, evidence of dynamic requirements has been observed according to the environments, economic, technology and manpower involved in the construction project. An adaptive design for project monitoring is needed which allow the physical system to self-adapt in response to the changing environments. Adaptive design requires selecting the right techniques of requirements elicitation. The conceptual framework defined shall be used to elicit requirements from a local construction company.

## 1. Introduction

Rapid change of technologies nowadays has become the reason why software systems are becoming inevitably more open, spread, persistent, mobile and connected [1]. For example, a mobile phone for the last 20 years in Malaysia may not has the internet access, but today the smart phone can provide the service for accessing the internet.

According to Alan Colman (2007), the inventor of ROAD framework, he said that this rapid changing environment requires the software system to interact with other software system and within heterogeneous and dynamic environments. Therefore, many types of software system need to cope with immediate changes in both requirements and environment and having to balance both these types of change [1]. Colman (2007) has approached ROAD as a meta-model and a framework for the construction of software applications that will be adaptive to both changing requirements and environments. Nevertheless, ROAD did not provide any methodology to be followed. Thus, the purpose of this paper is to highlight the methodology approach suggested for this research known as Participatory Action Research (PAR) and to implement it into the construction-based industry in Malaysia.

## 2. Requirement Engineering

Requirement Engineering (RE) plays a key role in software development in order to meet customer needs (Blanes.et.al, 2009). The main purpose of the requirements phase is to gather and elaborate a formal definition of the information needs the users have on the target collections [19]. RE includes eliciting, analyzing, validating and communicating stakeholder needs [14]. Many system engineers have been divided requirement levels into two categories which are high-level and low-level [14]. High-level requirements are described with words what customer requirements, top level requirements, system requirements, operational requirements, concept of operations, mission statement, stakeholder expectations, constraints, external requirements and what's [14].

While low-level requirements are described with words like derived requirements, product requirements, allocated requirements, internal requirements and how's [14]. Professor Eric Yu and Professor John Mylopoulos from University of Toronto had described the requirements providing the "why", design specifying the "what" and implementation giving the "how" [17]. In order to improve or redesign a process, it is important to have deeper understanding that reveals the "whys" behind the "whats" and the "hows" [19]. For a system to be successful, it must function within the context of its environment [18]. On 1997, Professor John Mylopoulos identified four main classes of modeling ontologies that are static, dynamic, intentional and social ontologies [18]. Static ontology describes static aspects of the world such as the entity-relationships models and class diagrams. A dynamic ontology describes the changing aspects of the world in terms of states, state transitions and processes.

## 3. The Famous System Failures

Various research studies have revealed that many errors can be introduced in a high-level model during the early phases of software requirements analysis and design. These errors can effects on reliability, cost, and safety of a software system [14]. Requirements errors come from two causes. Firstly, when software developers do not familiar with the application domain on which the software system is built. Secondly, they usually acquire requirements by talking to users. Misunderstandings between developers and users can result in requirements errors [14].

Table 1.1 below shows the twelve examples of famous system failures since 1912 which the main cause is in requirement development [14].

## 4. The Origin of PAR

PAR is an approach that originally proposed by an American psychologist, Kurt Lewin in the mid-1940s. Though, in some literature mentioned that the origins of action research are unclear, authors such as Kemmis and McTaggert (1988), Zuber-Skerrit (1992), Holter and Schwartz-Barcott (1993) stated that action research originated with Kurt Lewin [2]. McKernan (1991:8) also states that there is evidence of the use of action research by a number of social reformists prior to Lewin, such as Collier in 1945, Lippitt and Radke in 1946 and Corey in 1953 [2].

## 5. PAR Evolution

*"You cannot understand a system until you try to change it" (Lewin)*

Lewin once said to the world in describing the system which the author feel that it is very truthful reason why PAR should be celebrated in today development of the information system. At the early age of PAR exist in this world, Lewin named it as the Action Research. After several years, action research has gone through some evolution until Fals Borda (1970) came out with the importance of researchers to participate alongside with the participant in their research. Then, Action Research is known as Participatory Action Research [3].

| Name | Year | RD | VER | VAL |
|---|---|---|---|---|
| Most commercial systems | | Yes | Yes | Yes |
| Perpetual motion machine | | No | NB* | Yes |
| Titanic | 1912 | No | No | Yes |
| Tacoma Narrows Bridge | 1940 | Yes | Yes | No |
| Edsel automobile | 1958 | Yes | Yes | No |
| War in Vietnam | 1967-72 | Yes | Yes | No |
| Apollo-13 | 1970 | Yes | No | Yes |
| Concorde SST | 1976-2003 | Yes | Yes | No |
| IBM PCjr | 1983 | No | Yes | Yes |
| GE rotary compressor refrigerator | 1986 | Yes | No | Yes |
| Space Shuttle Challenger | 1986 | Yes | No | No |
| Chernobyl Nuclear Power Plant | 1986 | Yes | Yes | No |
| New Coke | 1988 | Yes | Yes | No |
| A-12 airplane | 1980s | No | NB* | No |
| Hubble Space Telescope | 1990 | Yes | No | Yes |
| SuperConducting SuperCollider | 1995 | Yes | Yes | No |
| Ariane 5 missile | 1996 | Yes | No | No |
| UNPROFOR Bosnia Mission | 1992-95 | No | No | No |
| Lewis Spacecraft | 1997 | Yes | Yes | No |
| Motorola Iridium System | 1999 | Yes | Yes | No |
| Mars Climate Orbiter | 1999 | No | No | No |
| Mars Polar Lander | 2000 | Yes | No | Yes |
| September 11 attack on WTT | 2001 | Yes | Yes | Yes |
| Space Shuttle Columbia | 2002 | Yes | No | No |
| Northeast power outage | 2003 | No | Yes | Yes |
| *NB means system was not built. | | | | |

*Table 2.1: Some famous system failures since 1912 until 2003* [14].

Action research is described by Lewin as a proceeding in a spiral of steps, which each of the steps composed of planning, action and the evaluation of the result of action [4]. Lewin argued that in order to understand and change certain social practices, social scientists have to include practitioners from the real social world in all phases of inquiry [2]. This construction of action research theory by Lewin made action research a method of acceptable inquiry [2].

In 1959, Carter identified four essential characteristics of action research [6]. There are:

i.  The problem for research must be generated from a recognized community need.

ii. The community members to be affected by the outcomes of research must be involved in the study process
iii. A team work approach amongst all involved in the research process is essential to effective outcomes.
iv. The research results must be in the form of recommendations for action or social change

In 1970, Orlando Fals-Borda, a Colombian sociologist, was able to effectively incorporate the "Community Action" component into the research plans of many traditionally trained researchers. It was not until then that communities started to fully appreciate the benefits of this approach which had initially seemed too abstract for many [2]. Fals-Borda highlighted that instead of the researcher needs to participate with the organization, researcher also needs to join the movement of change in the organization [7].

Early 20th century, Antonio Gramsci argues that all people are intellectuals and philophers. He defined it as "Organic intellectuals" where people who take their local knowledge from life experiences, and use that knowledge to address changes and problems in society. The idea that PAR researchers are really co-learners and researchers with the people they meet in the research process promotes the validity that all people are intellectuals who develop intricate philosophies through lived experience [8].

Supported by Greenwood (1993), PAR needs the full collaborators with the members of organizations in studying and transforming those organizations. It is an ongoing organizational learning process, a research approach that emphasizes co-learning, participation and organizational transformation [9].

Robin McTaggart (1997) in his book, *"Participatory Action Research: International Contexts and Consequences"*, described PAR as a broad church, movement, or family of activities where the movement expresses a recognition that all research methodologies are implicitly political in character, defining a relationship of advantage and power between the researcher and the researched. He also suggested that PAR required more than the validity of arguments to achieve acceptance by the research establishments it confronted and by the people it claimed to support [10].

Yoland Wadsworth (1998) also gives such contribution in defining for PAR. Wadsmorth argues on the research process where he described the important of the mix of its elements between action and participation. Action is important because all research is an action in itself and has consequences. PAR seeks to explicitly study something in order to change and improve the research study even though it does not start out with a precise idea. Wadsworth also proposed on the quality and depth of the theory and the design of the process to facilitate creativity [11]. Whilst participation is important as involvement between participant and researcher can reduce confusion or lack of agreement regarding the direction and purpose of the inquiry for whom and for what, improves the chances of asking the right questions, collecting the data and implementing changes. Wadsworth also argues why some action researcher neglect to use "participation" into the "action research". Wadsworth added with that participation, action and research are not separate in practice; it must come together into the research process [11].

In 2007, Paul Chatterton, Duncan fuller and Paul Routledge defined PAR as an approach that about jointly producing knowledge with others to produce critical interpretations and readings of the world, which are accessible, understandable to all those involved and actionable [11]. While Reason and Bradbury (2011) defined PAR as democratic process which concerned with developing practical knowing in the pursuit of worthwhile human purposes, grounded in a participatory worldview and bringing together action and reflection, theory and practice, in participation with others in the pursuit of practical solutions to issues of pressing concern to people, and more generally the flourishing of individual persons and communities [12].

## 6. PAR Process

Participatory action research (PAR) is a way of learning form and through one's practice by working through a series of reflective stages that facilitate the development of a form of "adaptive" expertise. Over time, action researchers develop a deep understanding of the ways in which a variety of social and environmental forces interact to create complex patterns. Since these forces are dynamic, action research is a process of living one's theory into practice [6].

The subject of action research is the actions taken, the resulting change, and the theory of change that is held by the persons enacting the change. While the design of action research may originate with an individual, social actions taken without the collaborative participation are often less effective. Over time, the action researchers often extend the arena of change to a continually widening group of stakeholders. The goal is a deeper understanding of the factors of change which result in positive personal and professional change.

This form of research then is an iterative, cyclical process of reflecting on practice, taking an action, observation, reflecting, and taking further action. Therefore, the research takes shape while it is being performed. Greater understanding from each cycle points the way to improved actions [6].

See Figure 1.1 below [13].

i. Addresses very clear, specific questions in a structured way.
ii. Integrates those stakeholders most affected by each particular question in the full process of research and decision making about this question.
iii. The agency proportionally allocates decision making power between stakeholders according to the level of effect of the answer to each question.

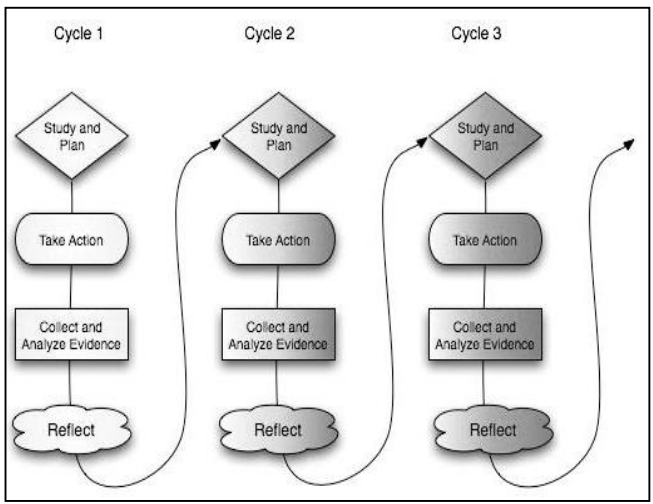

*Figure 1.1: Progressive Problem Solving with Action Research* [13].

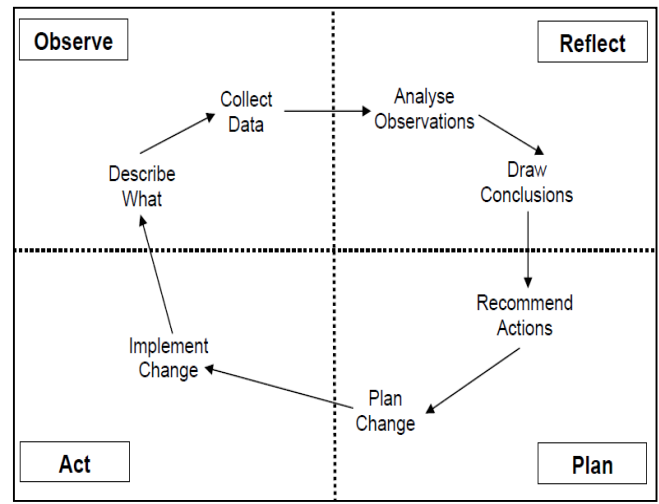

*Figure 1.2: Working PAR Model* [6]

The cycle process is dependable with the research problem and situation. A cycle does not necessarily begin with Planning. It could begin with any of the other processes - Acting, Observing or Reflecting. The process is a dynamic, constantly evolving one. Circles can overlap one another [6].

Figure 1.2 below described a working PAR model which consists of 4 phases [6].

A PAR Process enables specific stakeholders to own decisions about those aspects of service development and delivery that directly affect them. This is an improvement oriented model of practice, which is usually implemented by practitioners.

A PAR Process must consider about:

## 7. *i\* Framework*

*i\** (pronounced as "i star") or *i\** framework is a modelling language suitable for an early phase of system modelling in order to understand the problem domain. The name *i\** refers to the notion of distributed intentionality which underlines the framework. It is an approach originally developed for modelling and reasoning about organizational environments and their information systems composed of heterogeneous actors with different, often competing, goals that depend on each other to undertake their tasks and achieve these goals [15][16].

The *i\** modelling framework was attempt to bring in some aspects of social ontology which rarely respond in information system engineering compared to other ontologies [15][16]. The main concept in *i\** is about the intentionality of the actor. Actors are viewed as being intentional which means they have goals, belief, abilities, and commitments. Actors depend on each other for goals to be achieved, tasks and resources to be furnished. *i\** models offer a number of levels of analysis, in terms of ability, workability, viability and believability. It consists of two main modelling techniques which are Strategic Dependency (SD) model and Strategic Rationale (SR) model. SD model is used to describe the dependency relationships among various actors in organizational context. While SR model is used to describe stakeholder interests and concerns, and how might be addressed by various configurations of systems and environments [15][16].

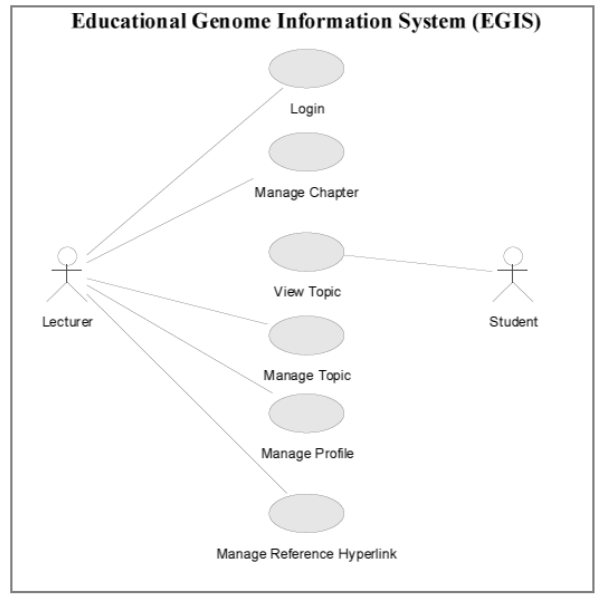

*Figure 1.4 : Use Case Diagram for EGIS*

Figure 1.3 below shows an example of Strategic Rationale (SR) Model using the implementation of *i\** modeling framework for educational genome information systems [20].

Table 1.2 shows the comparison between *i\**framework with UML.

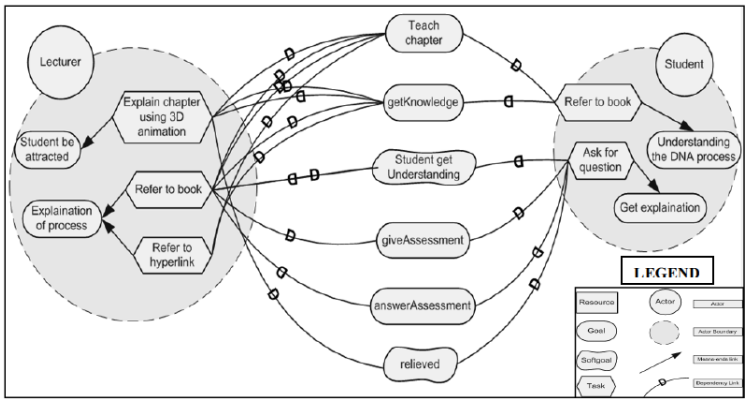

*Figure 1.3: SR Model of educational genome information system [20]*

After requirements are modeled using *i\** framework, then the researcher will continued to model it using UML. The resulted is as followed.

| *i\** Modeling | Unified Modeling Language |
|---|---|
| Use at *early* requirement phase | Use at the *late* requirement phase |
| Focus on *intentionality* of the actor | Focus on *operation* of the actor |
| Offer level of analysis including ability, workability, viability and believability | Cover only functional goals with actors directly involved in operation. |
| Answer WHO and WHY | Answer WHAT |

*Table 1.2 : Comparison between i\* Modeling Framework and UML [15][16]*

Since *i\** is used for early requirements and UML is used for late requirements, thus it needs to transform *i\** into a UML model. The following are the guidelines for transforming i\* into UML model.

    i.  *Actors* – Actors can be mapped to class aggregation.

...

ii. *Tasks* – Tasks can be mapped to class operations. For example, a task between a depender and a dependee actor in the SD model corresponds to a public operation in the dependee UML class.
iii. *Resources* – Resources can be mapped as classes
iv. *Goals* and *Soft goals* – Strategic goal and soft goals can be mapped to attributes
v. *Task decomposition* – Task decomposition can be represented by pre and post conditions.

## 8. Sequencing PAR and *i\** Framework in RE

Requirement engineering consists of several phases which are elicitation, analysis, specification and verification and validation. Since the methods of eliciting requirements are now more co-operative, this research proposed an idea on combining PAR methodology in capturing multiple roles requirements and understand the current situation of the research study alongside with the participants. After the requirements are captured, *i\** framework will take a place to analyze the requirements.

Figure 1.5 below shows a proposed model of sequencing PAR and *i\** framework in Requirement Engineering.

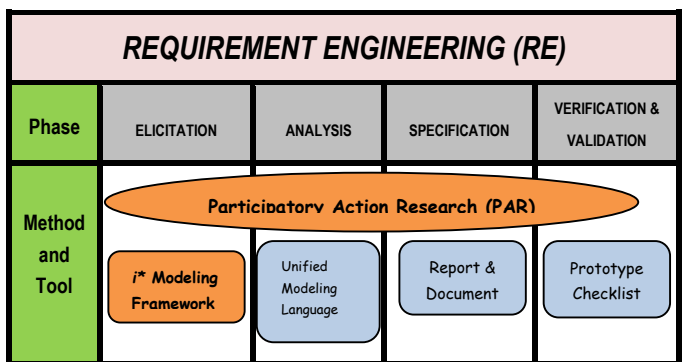

*Figure 1.5: A proposed model of sequencing PAR and i\* framework in Requirement Engineering*

## 9. A Brief Case Study

This research has implemented a proposed model stated above, to the one of Malaysian construction project in the end of 2012. The construction project is about the development of Malaysian new international airport hub that allows seamless connectivity for both local and international low-cost plus full services carriers known as Kuala Lumpur International Airport (KLIA) 2, which the scope project for this construction company is development for an airport runaway. Figure 1.6 shows the result from the implementation of the proposed framework. This research finds that by using PAR approach, researcher can get the horizontal overview of the business processes compared by using the use case driven approach where the researcher only gets the overview from the vertical side shows in Figure 1.7.

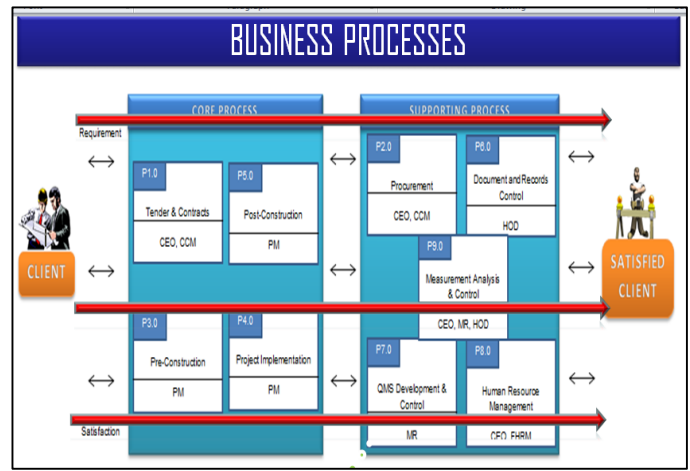

*Figure 1.6: Using PAR Approach - Horizontal overview of Business Process*

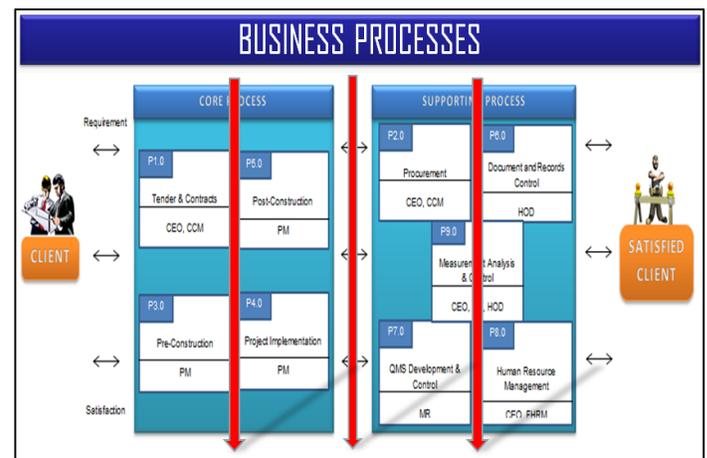

*Figure 1.7: Use case Driven Approach - Vertical overview of Business Process*

Result from this case study has been concluded in Figure 1.8 where using PAR approach, researcher can see the functional roles, process and sub-process for the case study in one shot of implementation, instead of using use-case driven approach where the researcher needs repetitive process to identify the functional roles to the process and later the sub-process.

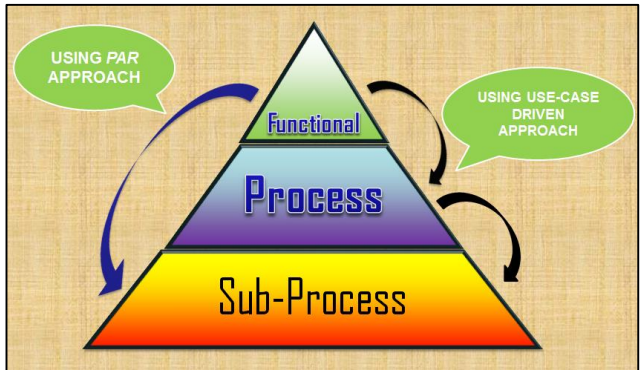

*Figure 1.8: Comparison between PAR Approach and Use Case Driven Approach*

## 10. Conclusion

In this paper, we have present the literature of PAR methodology from the different viewpoints of different authors starting from the problem of multiple roles involved in development of information system and the need of adaptive design in to be adapted into software system which nowadays becoming heterogeneous and dynamic environments and changing of requirements. We then proposed the participatory action research as a methodology to be followed together with adaptive design framework which is ROAD framework. We explained the origin as well as the evolution of PAR and its advantages. We further note the PAR process. PAR is in cyclical process which consists of four stages, planning, action, observing and reflecting. This research study wants to highlight that instead of using other conventionally technique in gathering user requirement, PAR is an interesting methodology to be celebrated in today development of information system.

This paper also presents the framework to analyse the complex system such as in construction industry which involved with multiple role-players as well as the brief description from the implementation of the proposed framework and the result from the case study.